# Superlight inverse Doppler effect

Xihang Shi[1†], Xiao Lin[1†*], Ido Kaminer[2*], Fei Gao[1], Zhaoju Yang[1], John D. Joannopoulos[2],

Marin Soljačić[2] & Baile Zhang[1,3*]

[1]Division of Physics and Applied Physics, School of Physical and Mathematical Sciences, Nanyang Technological University, Singapore 637371, Singapore.
[2]Department of Physics, Massachusetts Institute of Technology, Cambridge, MA 02139, USA.
[3]Centre for Disruptive Photonic Technologies, Nanyang Technological University, Singapore 637371, Singapore.
[†]These authors contributed equally to this work.
[*]Corresponding authors: xiaolinbnwj@ntu.edu.sg; kaminer@mit.edu; blzhang@ntu.edu.sg

**There is a century-old tenet [1, 2] that the *inverse* Doppler frequency shift of light [3-13] is impossible in *homogeneous* systems with a *positive* refractive index. Here we break this long-held tenet by predicting a new kind of Doppler effect of light inside the Cherenkov cone. Ever since the classic work of Ginzburg and Frank, it has been known that a *superlight* (i.e., superluminal) normal Doppler effect [14-18] appears inside the Cherenkov cone when the velocity of the source $v$ is larger than the phase velocity of light $v_\text{p}$. By further developing their theory we discover that an *inverse* Doppler frequency shift will arise when $v > 2v_\text{p}$. We denote this as the superlight *inverse* Doppler effect. Moreover, we show that the superlight inverse Doppler effect can be spatially separated from the other Doppler effects by using highly squeezed polaritons (such as graphene plasmons), which may facilitate the experimental observation.**

The Doppler effect, as a well-known phenomenon of motion-induced frequency shift, is one of the most fundamental mechanisms in physics and has vast applications in fields as varied as weather and aircraft radars, satellite global positioning systems, blood flow measurement in unborn fetal vessels, laser vibrometry, and the detection of extrasolar planets [1]. It occurs whenever the source and the observer move relative to each other. For the conventional Doppler frequency shift, the received frequency is higher (lower) compared to the emitted frequency during the approach (recession). In 1843, Christian Doppler propounded the conventional Doppler effect, first in relation to sound, and then to light [1, 2]. Since then, it is believed that the counterintuitive *inverse* Doppler frequency shift of light [2], where the sign of the frequency shift is opposite to that of the conventional Doppler frequency shift, cannot occur in homogeneous systems with a positive refractive index. In 1968, Victor Veselago predicted the inverse Doppler effect (with the inverse Doppler frequency shift) in systems with a negative refractive index [3-7]. Recently, more ways were proposed to realize the inverse Doppler effect, ranging from the use of shock waves [8-11] to the use of periodic structures [12, 13], but they all involved strongly inhomogeneous systems.

Here we find that it is indeed possible to create the inverse Doppler frequency shift in homogeneous systems with a positive refractive index inside the Cherenkov cone, as long as the radiation source moves with a velocity $v$ larger than twice the phase velocity $v_\text{p}$ of light, i.e., $v > 2v_\text{p}$. We denote this new phenomenon as the superlight *inverse* Doppler effect, which breaks the above century-old belief. The superlight inverse Doppler effect is different from the superlight normal Doppler effect [14-18], which, firstly studied by V. L. Ginzburg and I. M. Frank (GF) in 1947, emerges when $v > v_\text{p}$. While their theory predicted the superlight normal Doppler effect inside the Cherenkov cone, we reveal that the Doppler effects inside the Cherenkov cone can be divided into two categories, i.e., the superlight normal and superlight *inverse* Doppler effects. Therefore, our finding further develops the GF theory of the superlight normal Doppler effect.

For conceptual demonstration, we begin with the derivation of various Doppler effects of light. Consider that a radiation source (such as a point source with a dipole moment of $\bar{P}(\bar{r}',t') = Re\{\hat{x}e^{-i\omega_0 t'}\}\delta(\bar{r}'))$ moves in a system with a positive refractive index $n$ ($n > 0$) and has a natural angular frequency of $\omega_0$ ($\omega_0 > 0$) in the moving source frame. After applying the plane wave expansion [19], the frequency and wavevector in the two different frames (i.e., the lab frame and the moving source frame) can be directly linked through the Lorentz transformation [20], i.e.,

$$\begin{bmatrix} \bar{k} \\ \omega/c \end{bmatrix} = \begin{bmatrix} \bar{\bar{\alpha}} & +\gamma\bar{\beta} \\ +\gamma\bar{\beta} & \gamma \end{bmatrix} \begin{bmatrix} \bar{k}' \\ \omega'/c \end{bmatrix} \quad (1)$$

In equation (1), $\bar{k} = \hat{x}k_x + \hat{y}k_y + \hat{z}k_z$ ($\bar{k}' = \hat{x}k_x' + \hat{y}k_y' + \hat{z}k_z'$) and $\omega$ ($\omega' = \omega_0$) are the wavevector and the frequency in the lab frame (the moving source frame), respectively; $\bar{v} = +\hat{z}v$ is the velocity of the source, with its normalized form being $\bar{\beta} = \bar{v}/c$; $\gamma = (1-\beta^2)^{-1/2}$ is the Lorentz factor; finally, we use the definition $\bar{\bar{\alpha}} = \bar{\bar{I}} + (\gamma-1)\frac{\bar{\beta}\bar{\beta}}{\beta^2}$, with $\bar{\bar{I}}$ being the unity dyad. From the Lorentz transformation,

$$\omega = \gamma\omega_0 + \gamma v k_z' \quad (2)$$

$$k_z = \gamma\frac{v}{c}\frac{\omega_0}{c} + \gamma k_z' \quad (3)$$

In equation (2), $k_z' \in (-\infty, +\infty)$, and the positive (negative) value of $k_z'$ represents the generated waves propagating along the $+\hat{z}$ ($-\hat{z}$) direction in the source static frame. Since $\omega - \gamma\omega_0$ is proportional to $k_z'$, one also has $\omega \in (-\infty, +\infty)$, and has $\omega < 0$ ($\omega > 0$) when $k_z' < -\omega_0/v$ ($k_z' > -\omega_0/v$). By combining equations (2-3), $k_z = \frac{\omega - \omega_0/\gamma}{v}$ is derived. With the knowledge of $k_z = n\frac{\omega}{c}\cos\theta$, where $\theta$ is the angle between the velocity of the source and the wavevector $\bar{k}$ of the emitted photon (see Fig. 1), one can further express various Doppler effects of light in the following ordinary way

$$n\frac{v}{c}\omega\cos\theta = \omega - \omega_0/\gamma \tag{4}$$

Five radiation cases can be distinguished from equation (4), where the crucial factor that determines the type of the Doppler effect is the ratio between the velocity of the source and the z-component of the phase velocity ($v_p = c/n$) of light, i.e., $n\frac{v}{c}\cos\theta$.

*Case 1* - When $n\frac{v}{c}\cos\theta = 1$, equation (4) is valid only if $\omega_0 = 0$. This corresponds to the well-known Cherenkov radiation [20], where the Cherenkov cone is denoted in Fig. 1.

*Case 2* - When $n\frac{v}{c}\cos\theta < 1$, equation (4) is satisfied only for $\omega > 0$, and one has

$$\omega = \frac{\omega_0/\gamma}{1-n\frac{v}{c}\cos\theta}, \qquad n\frac{v}{c}\cos\theta < 1 \tag{5}$$

When the source moves towards (away from) the observer, i.e., $\theta < 90º$ ($\theta > 90º$), one has $\omega > \omega_0/\gamma$ ($\omega < \omega_0/\gamma$), where the appearance of $\gamma$ is due to the time dilation [20]. This corresponds to the conventional Doppler effect of light; see Fig. 1a.

When $n\frac{v}{c}\cos\theta > 1$, equation (4) is satisfied only for $\omega < 0$. As emphasized in the GF theory [14-16], having both frequencies $\omega_0$ and $\omega$ positive (see equation (5)) occurs if the source passes from an upper energy level to a lower one during the process of emission, i.e., the energy $\hbar\omega$ of the emitted photon is supplied from both the source excitation energy $\hbar\omega_0$ and its kinetic energy. However, having $\omega_0$ and $\omega$ of opposite signs (see equations (6,7) below) occurs if the source becomes excited by passing from a lower energy level to an upper one during the emission process, i.e., the kinetic energy of the source supplies both the energy of the emitted photon $\hbar|\omega|$ and the excitation energy $\hbar\omega_0$. We note that a similar transition between positive frequency and negative frequency has also been studied in other fields, such as negative refraction [21].

*Cases 3* - When $2 > n\frac{v}{c}\cos\theta > 1$, from equation (4), one has

$$|\omega| = \frac{\omega_0/\gamma}{n\frac{v}{c}\cos\theta - 1} > \omega_0/\gamma, \qquad 2 > n\frac{v}{c}\cos\theta > 1 \tag{6}$$

This is in accordance with the GF theory [14, 15] of the superlight normal Doppler effect (see Fig. 1b). The value of $|\omega|$ in the superlight normal Doppler effect increases with the value of $\theta$ and becomes infinity at the Cherenkov cone if the dispersion of refractive index $n$ is neglected. This is in stark contrast from the conventional Doppler effect, where the value of $\omega$ decreases with the value of $\theta$ (see equation (5)). Because of these differences from the conventional Doppler effect, the superlight normal Doppler effect is also denoted as the anomalous Doppler effect in some of the literature [16-18].

*Case 4* - When $n\frac{v}{c}\cos\theta = 2$, from equation (4), this leads simply to $|\omega| = \omega_0/\gamma$.

*Case 5* - When $n\frac{v}{c}\cos\theta > 2$, from equation (4), one has

$$|\omega| = \frac{\omega_0/\gamma}{n\frac{v}{c}\cos\theta - 1} < \omega_0/\gamma, \qquad n\frac{v}{c}\cos\theta > 2 \tag{7}$$

This corresponds to the superlight *inverse* Doppler effect (see Fig. 1c). The condition for its occurrence is $v > 2c/n$ and the value of $|\omega|$ increases with the value of $\theta$. The superlight inverse Doppler effect in equation (7) is different from the superlight normal Doppler effect in equation (6), where the former has $|\omega| < \omega_0/\gamma$ while the latter has $|\omega| > \omega_0/\gamma$. Therefore, equations (6-7) describe two different types of Doppler frequency shifts inside the Cherenkov cone. The superlight

inverse Doppler effect is a new Doppler phenomenon of light, and to our knowledge has never been discussed before. In addition, since the condition of $n\frac{v}{c}\cos\theta > 1$ is possible only for $\theta < 90°$, one always has $\theta < 90°$ for the superlight normal and superlight inverse Doppler effects; see Fig. 1.

Figure 2 schematically shows the difference between the different Doppler effects in the time domain, illustrated by a cross section cut in space, marking multiple phase-fronts that are equally-distributed in time. For the simplicity of conceptual demonstration, Fig. 2 also notes the Doppler-shifted wavelength along $\theta = 0°$ (i.e., $\bar{k}$ and $\bar{v}$ are in the same direction). There, both the conventional and superlight normal Doppler effects (i.e., when $v < 2c/n$) have the distance between successive wave fronts reduced, so the waves bunch together (Fig. 2a-e); in contrast, the superlight inverse Doppler effect (i.e., when $v > 2c/n$) has the distance between successive wave fronts enlarged, so the waves spread out ( Fig. 2f). This shows a clear difference between the superlight inverse Doppler effect and the two other kinds of Doppler effect. Figure 2 also highlights a clear difference between the conventional Doppler effect and the two superlight Doppler effects: at $\theta = 0°$, for the conventional Doppler effect (i.e., when $v < c/n$), the observer first receives the wave front emitted at an earlier time (Fig. 2b); however, for the superlight normal and superlight inverse Doppler effects (i.e., when $v > c/n$), the observer first receives the wave front emitted at a later time (Fig. 2d-f), corresponding to $\omega < 0$ in equations (6,7).

In order to illustrate the occurrence of the superlight inverse Doppler effect, figure 3 shows various Doppler effects of highly squeezed polaritons. The high effective refractive index of such polaritons [22-24] enables the superlight inverse Doppler effect to occur with a small value of $v$ (due to the condition of its occurrence $v > 2c/n$). As an example, Fig. 3 uses the surface plasmon polaritons (SPP) in graphene to demonstrate the superlight inverse Doppler effect at a small value of $v$. Assume that the source is a dipole with $\omega_0/2\pi = 10$ THz, moving parallel to a graphene monolayer; the graphene has a chemical potential of 0.15 eV and a relaxation time of 0.3 ps. Then graphene plasmons have an effective refractive index $n = \frac{k_{spp}}{|\omega|/c} = 19$ at $|\omega|/2\pi = 10$ THz, where $\bar{k}_{spp} = \hat{x}k_x + \hat{z}k_z$ ($k_{spp} = |\bar{k}_{spp}|$) is the plasmonic wavevector parallel to the graphene plane. Using this wavevector, all the Doppler effects for graphene plasmons can be described by a single equation [25]

$$|\omega| \pm \omega_0/\gamma = k_{spp} v \cos\theta \qquad (8)$$

where $\theta$ is the angle between $\bar{v}$ and $\bar{k}_{spp}$. When the minus (plus) sign is adopted, equation (8) characterizes the conventional (superlight normal and superlight inverse) Doppler effect of graphene plasmons, similar to equation (5) (equations (6,7)). When $v = 0.1c$, one has $2v_{p0} > v > v_{p0}$ ($v_{p0} = c/n = c/19$), and thus there is only the conventional and superlight normal Doppler effects; see dashed lines in Fig. 3. When $v = 0.3c$, one has $v > 2v_{p0}$; this way, the superlight inverse Doppler effect also emerges, as shown by the solid lines in Fig. 3 that cross between the two regimes. (We have considered the frequency dispersion of graphene plasmons [25] for all the results in Figs 3, 4.)

The experimental observation of the superlight inverse Doppler effect from a moving source (such as that in Fig. 3) can benefit from the detection of the angular (i.e., $\theta$-resolved) and spectral ($\omega$-resolved) energy density of the radiation, due to the dependence of the Doppler frequency shift on the radiation angle $\theta$. This may be extremely complex. In order to facilitate the observation, it is thus beneficial to reduce this complexity with the design of a novel scheme that enables us to judge the type of the Doppler effect without the need to detect the radiation angle $\theta$. Such a scheme is proposed in Fig. 4.

Figure 4 computationally shows that it is possible to spatially separate the superlight inverse Doppler effect through the judicious design of the moving source, which might foster the future observation. Consider a circularly-polarized source, which moves along the $+\hat{z}$ direction and has a dipole moment of $\bar{P}(\bar{r}', t') = Re\{(\hat{x}p_x + \hat{y}ip_y)e^{-i\omega_0 t'}\}\delta(\bar{r}')$ with $p_x = p_y = 1$ in the source static frame. Such circularly-polarized sources have been widely exploited for asymmetric excitation when $v = 0$ [26]. Figure 4a shows the distribution of the emitted plasmonic field in the graphene plane in the time domain [25]; see the dynamics of the emitted plasmons in Supplementary Video. When $v = 0.3c$, two asymmetric caustics [27-29] are formed in the regions $x < 0$ (the left side of the source) and $x > 0$ (the right side of the source), respectively. Since the caustic frequency is close to the frequency of the wave component that dominates the plasmonic emission at each caustic [27, 28], it can be used to determine the type of Doppler effect that dominates each region. Due to the asymmetry of the two caustics in Fig. 4a, regions $x < 0$ and $x > 0$ have different caustic frequencies. The caustic frequency is calculated from $\frac{d^2\varphi}{d\omega^2} = 0$ [25, 27, 28], where $\varphi = k_x \frac{x}{t} + k_z \frac{z}{t} - \omega$ [25].

Figure 4b, c shows the two caustic frequencies $\omega_{\text{caustic}}$ and the progagation angles $\theta_{\text{caustic}}$ (the anlge between $\bar{k}_{\text{spp}}$ and $\bar{v}$ at each caustic frequency) for the regions $x < 0$ and $x > 0$ as a function of $v$, respectively. From Fig. 4b, c, one can see that the region $x < 0$ is dominated by the conventional Doppler effect with $\omega_{\text{caustic}} < \omega_0/\gamma$, independent of the value of $v$. In contrast, the region $x > 0$ is dominated by the superlight normal Doppler effect with $\omega_{\text{caustic}} > \omega_0/\gamma$ when $v < 0.15c$, and, importantly, becomes dominated by the superlight inverse Doppler

effect with $\omega_{\text{caustic}} < \omega_0/\gamma$ when $v > 0.15c$. It shall be emphasized that, when $v > 0.15c$, the excited plasmon in region $x < 0$ (having the conventional Doppler effect) propagates along the backward ($-\hat{z}$) direction with $\theta_{\text{caustic}} > 90^o$; in contrast, the excited plasmon in region $x > 0$ (dominant by the superlight inverse Doppler effect) propagates along the forward ($+\hat{z}$) direction with $\theta_{\text{caustic}} < 90^o$; see Fig. 4a and Supplementary Video for an example.

To conclude, we further develop the GF theory of superlight normal Doppler effect by predicting that even inside the Cherenkov cone, the Doppler effects of light can be divided into two categories, i.e., the superlight normal and superlight inverse Doppler effects. Most importantly, the revealed superlight inverse Doppler effect breaks the century-old tenet that the inverse Doppler frequency shift is impossible in homogeneous systems with a positive refractive index. Moreover, we have numerically demonstrated the possibility for the spatial separation of the superlight inverse Doppler effect from the other Doppler effects, thus facilitating the detection of this new phenomenon. Perhaps even more important is the vision our findings emphasize: that the analogous phenomena of the superlight inverse Doppler effect will exist in virtually any wave system in nature, including classical wave systems such as acoustic waves and surface waves, as well as quantum wave systems such as the Dirac equation; all of which have shown the occurrence of the Doppler effect, and therefore can now also support the analogous phenomena of the superlight inverse Doppler effect. We note that the inverse Doppler effect of acoustic waves was experimentally observed in acoustic metamaterials and phononic crystals recently [30, 31]. Therefore, in additional to the highly squeezed polaritons, the acoustic waves with a slow phase velocity may provide another promising platform for the experimental demonstration of this new revealed Doppler effect.

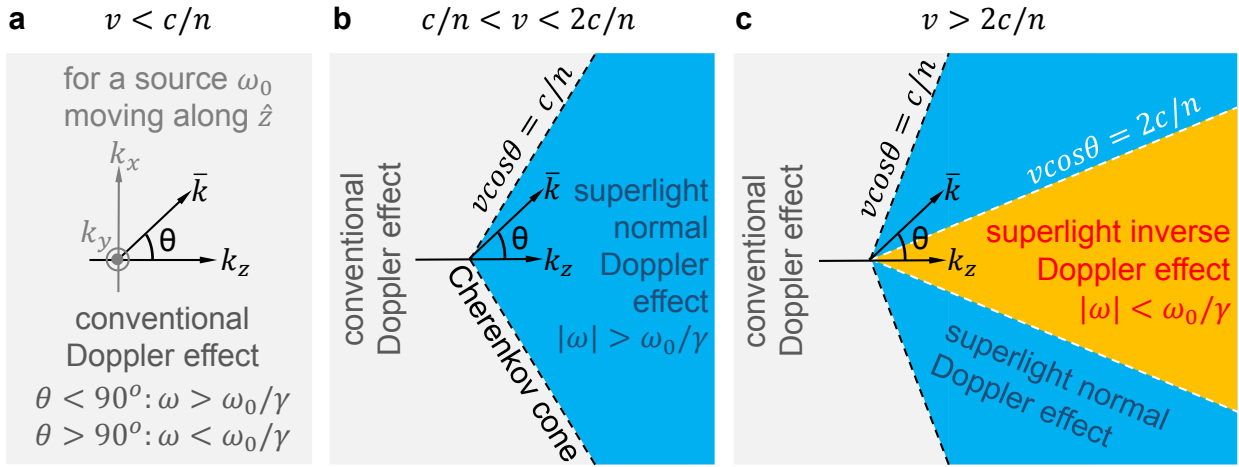

**Figure 1 | K-space representation of the conventional Dopper effect, superlight normal Doppler effect and superlight inverse Doppler effect.** A radiation source moves with a velocity of $\bar{v}$ along the $\hat{z}$ direction in a homogeneous system with a refractive index $n$ ($n > 0$). **a**, Only the conventional Dopper effect exists when $v < c/n$. **b**, When $v > c/n$, there is only the superlight normal Doppler effect inside the Cherenkov cone. **c**, When $v > 2c/n$, the superlight inverse Doppler effect also appears inside the Cherenkov cone. In the source static frame, the source has a natural frequency $\omega_0$. In the observer static frame, the received radiation fields have a frequency $\omega$ and a wavector $\bar{k}$, where $\theta$ is the angle between $\bar{k}$ and $k_z$ axis (or $\bar{v}$). $\gamma$ is the Lorentz factor. The Cherenkov cone is determined by the condition of $v\cos\theta = c/n$.

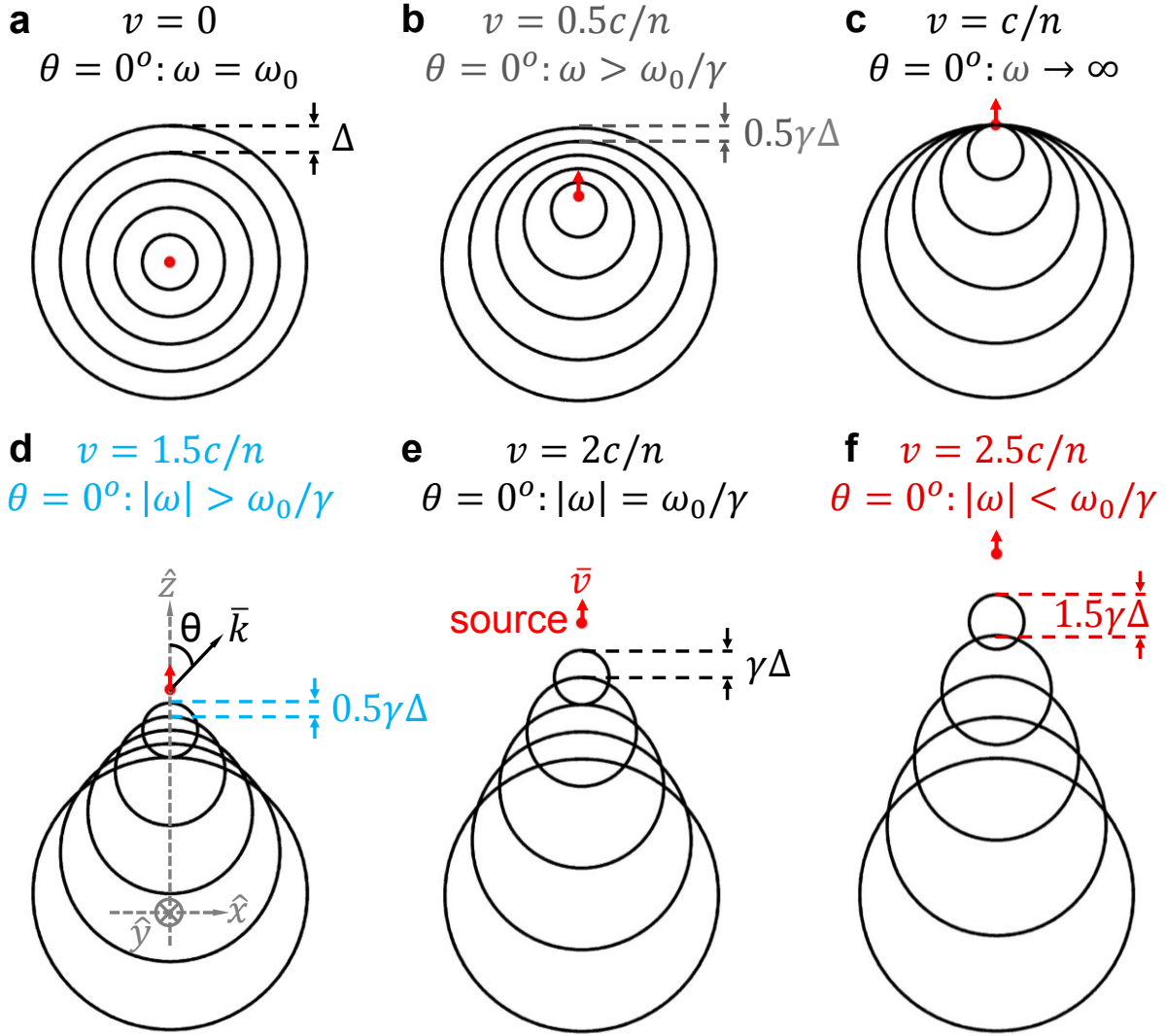

**Figure 2 | Real-space schematic demonstration of various Doppler effects of light.** The radiation source (represented by the red dot) moves along $\hat{z}$ direction in a homogeneous system with a constant positive refractive index ($n > 2.5$). The multiple phase-fronts, illustrated by circular lines, are equally-distributed in time. $\theta$ is the angle between $\bar{k}$ and $\hat{z}$ (or $\bar{v}$); when $\theta = 0^o$, $\bar{k}$ and $\hat{z}$ are in the same direction. **a**, When $v = 0$, there is no Doppler effect. **b**, When $v < c/n$, the conventional Doppler effect exists at $\theta = 0^o$. **c**, When $v = c/n$, the value of $|\omega|$ goes to infinity at $\theta = 0^o$. **d**, When $c/n < v < 2c/n$, the superlight normal Doppler effect appears at $\theta = 0^o$. When $v < 2c/n$, the distance between successive wave fronts at $\theta = 0^o$ (labelled in each

panel) is reduced and the waves bunch together. This leads to $|\omega| > \omega_0/\gamma$ at $\theta = 0°$ in (b-d). **e**, When $v = 2c/n$, $|\omega| = \omega_0/\gamma$ at $\theta = 0°$. **f**, When $v > 2c/n$, the distance between successive wave fronts at $\theta = 0°$ is enlarged and the waves spread out. This leads to the superlight inverse Doppler effect ($|\omega| < \omega_0/\gamma$) at $\theta = 0°$ in (f). The Lorentz factor $\gamma$ is different in each panel.

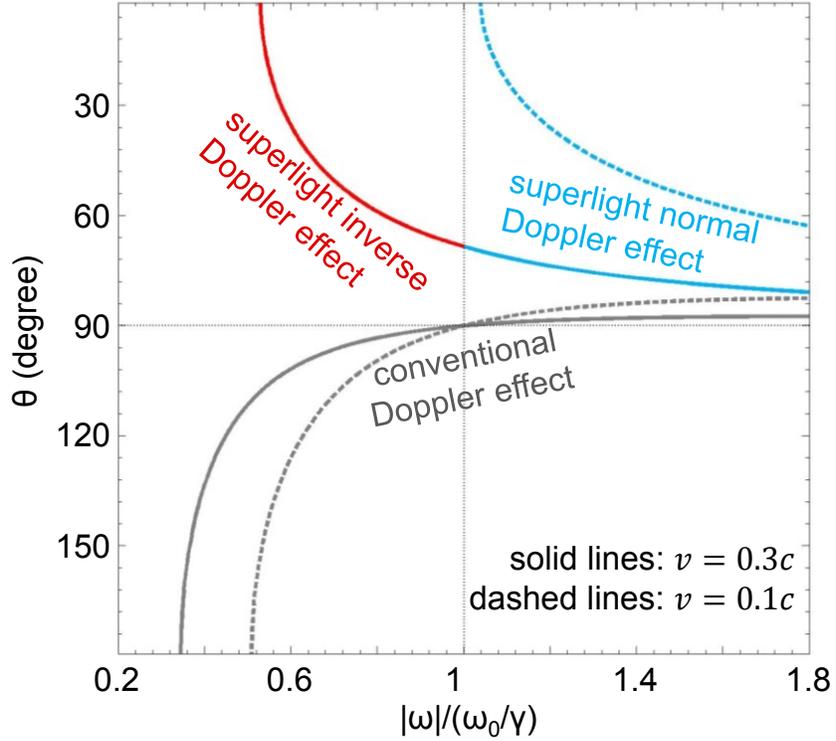

**Figure 3 | Superlight inverse Doppler effect of graphene plasmons.** A dipole with $\omega_0/2\pi = 10$ THz moves parallel to a graphene monolayer surrounded by air. When $v = 0.1c$, only the conventional and superlight normal doppler effects exist. When $v = 0.3c$, the superlight inverse doppler effect also occurs. The chemical potential of graphene is 0.15 eV and the relaxation time is 0.3 ps. Here, and in the figure below, the frequency dispersion of graphene plasmons is considered and the effective refractive index of graphene plasmons is $n = 19$ at 10 THz. The superlight inverse Doppler effect thus appears only when $v > 2c/n = 2c/19$.

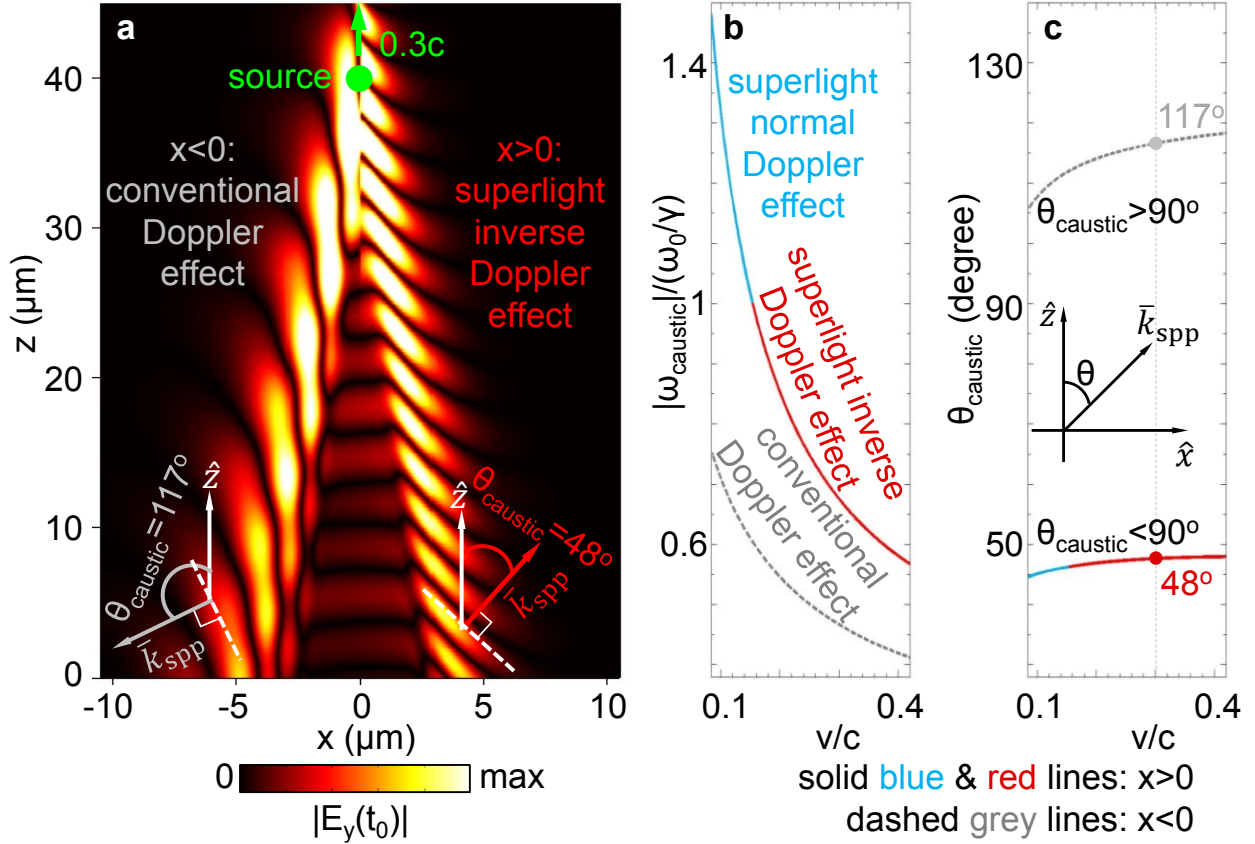

**Figure 4 | Real-space representation of the superlight inverse Doppler effect spatially separated from other Doppler effects.** A circularly-polarized dipole source with $\omega_0/2\pi = 10$ THz moves along the $+\hat{z}$ direction and parallel to the graphene monolayer. **a**, Distribution of the asymmetric radiation fields on the graphene plane with $v = 0.3c$ at the fixed time $t = t_0$ (when the source is at $z = 40$ μm). The excited plasmons propagate backward in the region $x < 0$ (the left side of source), while they propagate forward in the region $x > 0$ (the right side); see the direction of the plasmonic wavevector $\bar{k}_{\text{spp}}$ in the inset. **b**, The two caustic frequencies $|\omega_{\text{caustic}}|$ as a function of $v$; these two frequencies dominate the plasmon emission in the regions $x < 0$ and $x > 0$, respectively. **c**, Progagation angles $\theta_{\text{caustic}}$, i.e., the angle between $\bar{k}_{\text{spp}}$ and $\hat{z}$ (see inset) at each caustic frequency. When $v = 0.3c$, $\theta_{\text{caustic}}$ is schematically shown in (a). The region $x > 0$ is dominated by the superlight inverse Doppler effect when $v > 0.15c$ (see (a) for example). The

graphene (surrounded by air and parallel to the *x-z* plane) has a chemical potential of 0.15 eV and a relaxation time of 0.3 ps; the vertical distance between graphene and the source is 1 μm.